\documentclass[journal,transmag]{IEEEtran}
\usepackage{cite}
  \usepackage[pdftex]{graphicx}
\usepackage{amsmath}

\newcommand{\m}{\mathbf{m}}
\newcommand{\h}{\mathbf{h}}
\newcommand{\kb}{\mathbf{k}}
\newcommand{\ub}{\mathbf{u}}
\newcommand{\q}{\mathbf{q}}

\newcommand{\ubar}{\bar{u}}

\newcommand{\pp}[2]{\frac{\partial #1}{\partial #2}}

\begin{document}
%
\title{Magnonic Band Structure Established by Chiral Spin-Density Waves in Thin Film Ferromagnets}


\author{\IEEEauthorblockN{Patrick Sprenger\IEEEauthorrefmark{1},
Mark A. Hoefer\IEEEauthorrefmark{1}, and 
Ezio Iacocca\IEEEauthorrefmark{1}}
\IEEEauthorblockA{\IEEEauthorrefmark{1}Department of Applied Mathematics, University of Colorado Boulder, Boulder, CO 80302, USA}
}



\IEEEtitleabstractindextext{%
\begin{abstract}
 
Recent theoretical studies have demonstrated the possibility to excite and sustain noncollinear magnetization states in ferromagnetic nanowires. The resulting state is referred to as a spin-density wave (SDW). SDWs can be interpreted as hydrodynamic states with a constant fluid density and fluid velocity in systems with easy-plane anisotropy. Here, we consider the effect of the nonlocal dipole field arising from the finite thickness of magnetic thin films on the spatial profile of the SDW and on the associated magnon dispersion. Utilizing a hydrodynamic formulation of the Larmor torque equation, it is found that the nonlocal dipole field modulates the fluid velocity. Such a modulation induces a magnonic band structure unlike the typical dispersion relation for magnons on uniform magnetization. The analytical results are validated by micromagnetic simulations. Band gaps on the order of GHz are numerically observed to depend on the SDW fluid velocity and film thickness for realistic material parameters. The presented results suggest that SDWs can find applications as reconfigurable magnonic crystals.

\end{abstract}
}
\maketitle

\IEEEdisplaynontitleabstractindextext

\IEEEpeerreviewmaketitle

\section{Introduction}

Recent technological advances rely on magnetic materials for information transport, storage, and logic applications~\cite{Roadmap2017}. These functions can be realized via spin waves, the fundamental magnetic excitation in magnetic systems. From a quantum-mechanical perspective, spin waves are associated with bosonic quasiparticles known as magnons~\cite{White2007}. The field that studies and controls the dispersion of magnons is called \textit{magnonics}~\cite{Demokritov2013,Neusser2009}. By patterning magnetic super-lattices or magnonic crystals, magnonic band structures and band gaps~\cite{Krawczyk2014} that can be actively reconfigured~\cite{Grundler2015,han2009magnetic} are achieved. However, reconfigurable magnonic crystals are still challenging to realize because of the need to modify the energy landscape~\cite{Karenowska2012,Wang2017} or the long-range magnetic order in the crystal~\cite{Tacchi2011,Jungfleisch2016,Iacocca2016,Iacocca2017c,Gubbiotti2018}.

An alternative route to realize magnonic crystals may be found in noncollinear magnetization states. In contrast to patterned super-lattices, the periodicity of such states is determined by the material's magnetic parameters, e.g., in skyrmion lattices~\cite{Garst2017}. Ferromagnetic materials with easy-plane anisotropy can also host periodic textures known as spin-density waves (SDWs)~\cite{Iacocca2017} or spin superfluids~\cite{Konig2001}. While SDWs are theoretical constructs for infinitely extended, zero-thickness films, recent numerical and analytical studies have shown that spin injection can sustain noncollinear states in nanowires~\cite{Sonin2010,Takei2014,Iacocca2017b,Iacocca2017d}. However, the effect of the nonlocal dipole field arising from the finite film thickness on SDWs is still an open question. In this letter, we utilize a dispersive hydrodynamic model that includes weak nonlocal dipole fields due to the small but finite film thickness to describe symmetry-broken SDWs and determine the associated magnon dispersion.

It is found that symmetry-broken SDWs give rise to a magnonic band structure in the absence of an external magnetic field. Interestingly, the first Brillouin zone is given by \emph{twice} the SDW wavenumber. In the ideal case of a zero-thickness film, a band structure is not observed despite the SDW periodicity. These results suggest that noncollinear states in ferromagnets may serve as reconfigurable magnonic crystals without requiring physical patterning or active engineering of the energetic landscapes.

\section{Dispersive Hydrodynamic Model}

The magnetization dynamics in an ideal conservative ferromagnet are described by the Larmor torque equation
\begin{equation}\label{eq:torque}
\frac{\partial \mathbf{M}}{\partial t} = - \gamma\mu_0\mathbf{M} \times \mathbf{H}_{\rm eff}, 
\end{equation}
where $\mathbf{M} = [M_x,M_y,M_z]$ is the magnetization vector, $\mathbf{H}_{\rm eff}$ is an effective field, $\gamma$ is the gyromagnetic ratio, and $\mu_0$ is the vacuum permeability. A minimal model for a ferromagnetic thin film includes the effect of exchange interaction in a micromagnetic approximation and nonlocal dipole field, which yields the effective field
\begin{equation}\label{eq:heff}
\mathbf{H}_{\rm eff} = \lambda_\mathrm{ex}^2\Delta \mathbf{M} + \mathbf{H}_{\rm d},
\end{equation}
where $\lambda_\mathrm{ex}$ is the exchange length and $\mathbf{H}_{\rm d}$ is the nonlocal dipole field arising from the thickness $d$.

It is convenient to rescale the temporal and spatial scales by 
\begin{equation}\label{eq:nondim}
  t \to \gamma\mu_0 M_s t', \quad x \to  \lambda_\mathrm{ex}x'
  \end{equation} and fields by $M_s$, where $M_s=|\mathbf{M}|$ is the saturation magnetization. Inserting the dimensionless variables  and dropping primes, we arrive at
\begin{equation}\label{eq:torque2}
\frac{\partial \m}{\partial t} = - \m \times \left[\Delta \m + \h_{\rm d}\right], 
\end{equation}

We consider a thin film of dimensionless thickness $\delta=d/\lambda_\mathrm{ex}$. In the thin film regime $0 < \delta \ll 1$ the nonlocal dipole field can be approximated as~\cite{GarciaCervera2004,Bookman2013}
\begin{align}\label{eq:nl_dipole}
\h_{\rm d} = -m_z \hat{\mathbf{z}} + \delta \mathcal{F}^{-1}\left\{ \frac{-\kb (\kb \cdot \mathcal{F}\left\{\m_{\|}\right\})}{2|\kb|}- \frac{|\kb|\mathcal{F}\left\{m_z\right\}}{2} \hat{ \bf z} \right\},
\end{align}
where $\mathcal{F}$ and $\mathcal{F}^{-1}$ are respectively the forward and inverse Fourier transforms, $\kb = [k_x,k_y]$ is the wavevector, and $\m_\|=[m_x,m_y]$. The magnetization is assumed to be uniform through the thickness, $\mathbf{m}=\mathbf{m}(x,y,t)$. For zero-thickness, $\delta = 0$, Eq.~\eqref{eq:nl_dipole} reduces to the usual local dipole field approximation, $\mathbf{h}_d=-m_z\hat{\mathbf{z}}$.

A dispersive hydrodynamic model for Eq.~\eqref{eq:torque2} can be found 
through the canonical Hamiltonian variable transformation
\begin{align}
\label{eq:transf_n}
n &= m_z,\\
\label{eq:transf_u}
\ub &= - \nabla \phi  = - \nabla \arctan \frac{m_y}{m_x}, 
\end{align}
where $\phi$ is the phase of the in-plane magnetization. The variable transformation exactly reduces the vector partial differential equation \eqref{eq:torque2} to a system of two nonlinear dispersive partial differential equations 
\begin{align}
\label{eq:hydro_n}
\begin{split}
\pp{n}{t} & = \nabla \cdot \left[(1-n^2) \ub\right]\\& \quad + \sqrt{1-n^2} \left[\h_{\rm d} \cdot \hat{\bf x} \sin \phi - \h_{\rm d} \cdot \hat{\bf y } \cos \phi\right],
\end{split} \\
\begin{split}
\label{eq:hydro_phi}
\pp{\phi}{t} & = \h_{\rm d}\cdot\hat{\mathbf{z}} +|\mathbf{u}|^2n  + \frac{\Delta n}{1-n^2} + \frac{n |\nabla n|^2}{(1-n^2)^2}\\ & \quad  - \frac{n}{\sqrt{1-n^2}} \left[\h_{\rm d} \cdot   \hat{\bf x} \cos \phi + \h_{\rm d} \cdot   \hat{\bf y} \sin \phi \right]. 
\end{split}
\end{align}

Due to the fluid-like form of Eqs.~\eqref{eq:hydro_n} and \eqref{eq:hydro_phi}, we refer to $n$ as the signed \emph{fluid density} and to $\ub$ as the \emph{fluid velocity}. Because the magnetization vector is bounded in magnitude, the fluid density is likewise bounded as $|n|\leq1$. The fluid velocity is, in principle, arbitrary, but $|\mathbf{u}|>1$ coincides with length scales shorter than the exchange length. 

\section{Spin-density wave solutions}

\begin{figure}[t]
\begin{center}
\includegraphics[scale=0.25]{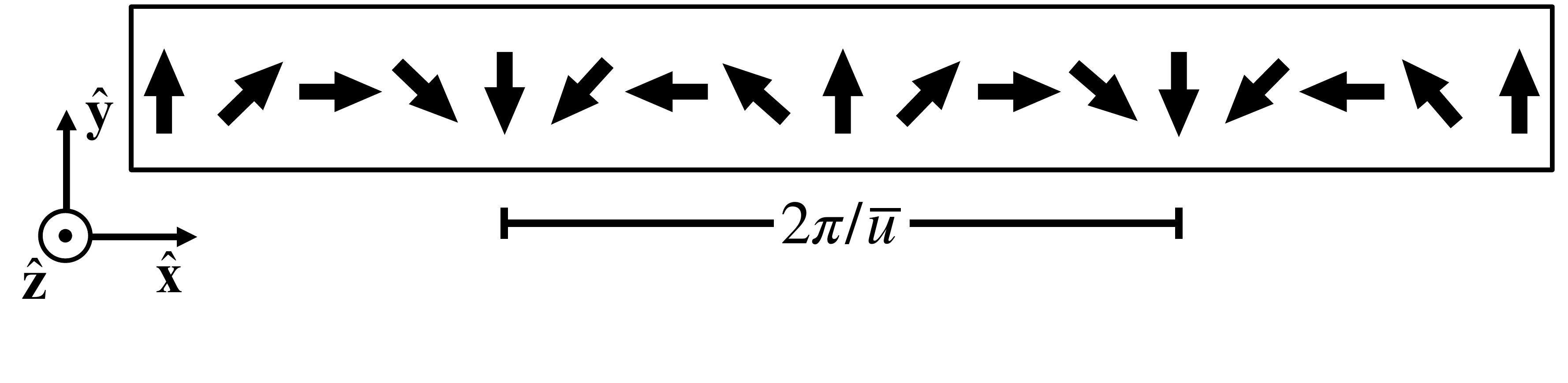}
\end{center}
\caption{\label{fig:SDW_schematic} Top view schematic of a spin density wave in a planar ferromagnet with velocity $\ub = \ubar \hat{\mathbf{x}}$. }
\end{figure}

In an ideal infinite planar ferromagnet with zero thickness, the nonlocal dipole field in Eq.~\eqref{eq:nl_dipole} reduces to $\mathbf{h}_{\rm d}=-n\hat{\mathbf{z}}$. Inserting this approximation into Eqs.~\eqref{eq:hydro_n} and \eqref{eq:hydro_phi} we can find a family of constant, static hydrodynamic solutions~\cite{Iacocca2017,Iacocca2017b}
\begin{align}\label{eq:SDW}
n & = 0, \quad
\ub  = \ubar \hat{\bf x}.,
\end{align}
each of which represents a SDW with wavelength $\lambda=2\pi/\ubar$. A schematic of a SDW is shown in Fig.~\ref{fig:SDW_schematic}. It has been shown that SDWs are stable for $|\ubar| < 1$~\cite{Iacocca2017}. This corresponds to SDW physical wavelengths larger than $\lambda_\mathrm{ex}$. SDWs with shorter wavelengths are modulationally unstable in the micromagnetic approximation~\cite{Iacocca2017}.

When a finite thickness is considered, the SDW~\eqref{eq:SDW} is no longer a solution to Eqs.~\eqref{eq:hydro_n} and \eqref{eq:hydro_phi}. In the thin film limit, an approximate SDW solution can be found by an asymptotic procedure. For this, we seek a correction in the hydrodynamic variables. Using the nondimensional thickness as a small parameter, $\delta$, we perform the asymptotic expansions
\begin{subequations}\label{eq:asymp}
\begin{eqnarray}
n &=& \delta n_1 + \mathcal{O}(\delta^2), \\
\ub &=& \ubar \hat{\bf x} + \delta \ub_1 + \mathcal{O}(\delta^2),\\
\phi &=& -\ubar x + \delta \phi_1 + \mathcal{O}(\delta^2),
\end{eqnarray}
\end{subequations}
where $\mathcal{O}(\delta^2)$ indicates corrections of order $\delta^2$ or higher. Evaluating Eq.~\eqref{eq:nl_dipole} on the SDW expansion Eq.~\eqref{eq:asymp} yields the approximate dipole field accurate to terms proportional to $\delta$,
\begin{equation}\label{eq:nlcorr}
\h_{\rm d} = -\delta n_1\hat{\mathbf{z}}+\frac{\delta |\ubar|}{2}\cos \left(\ubar x \right) \hat{\bf x}+ \mathcal{O}(\delta^2). 
\end{equation}

Inserting the expansion \eqref{eq:asymp} and the nonlocal dipole field \eqref{eq:nlcorr} into Eqs.~\eqref{eq:hydro_n} and \eqref{eq:hydro_phi}, the hydrodynamic equations at $O(\delta)$ are
\begin{subequations}
\begin{align}\label{eq:linearized1}
\pp{n_1}{t} & = \nabla \cdot \ub _1+ \frac{\ubar}{2}\cos \ubar x \sin \ubar x,\\
\pp{\phi_1}{t} & = \left(1 - \ubar^2\right)n_1 + \Delta n_1. \label{eq:linearized2}
\end{align}
\end{subequations}
Equations~\eqref{eq:linearized1} and \eqref{eq:linearized2} admit the steady, in-plane solution $n_1=0$ and $\mathbf{u}_1=(\delta/8)\cos{(2\bar{u}x)}\hat{\mathbf{x}}$, so that the approximate solution for a SDW in a finite thickness ferromagnetic film is
\begin{subequations}\label{eq:sdw_pert}
\begin{eqnarray}
\label{eq:nasymp}
n &\sim& \mathcal{O}(\delta^2), \\
\label{eq:uasymp}
\mathbf{u} &\sim& \ubar\hat{\mathbf{x}}  + \frac{\delta}{8} \cos (2 \ubar x)\hat{\mathbf{x}} +\mathcal{O}(\delta^2),\\
\phi & \sim& -\ubar x - \frac{\delta}{16}\sin( 2\ubar x) + \phi_0+\mathcal{O}(\delta^2),
\end{eqnarray}
\end{subequations}

\begin{figure}[t]
\begin{center}
\includegraphics[scale=0.3]{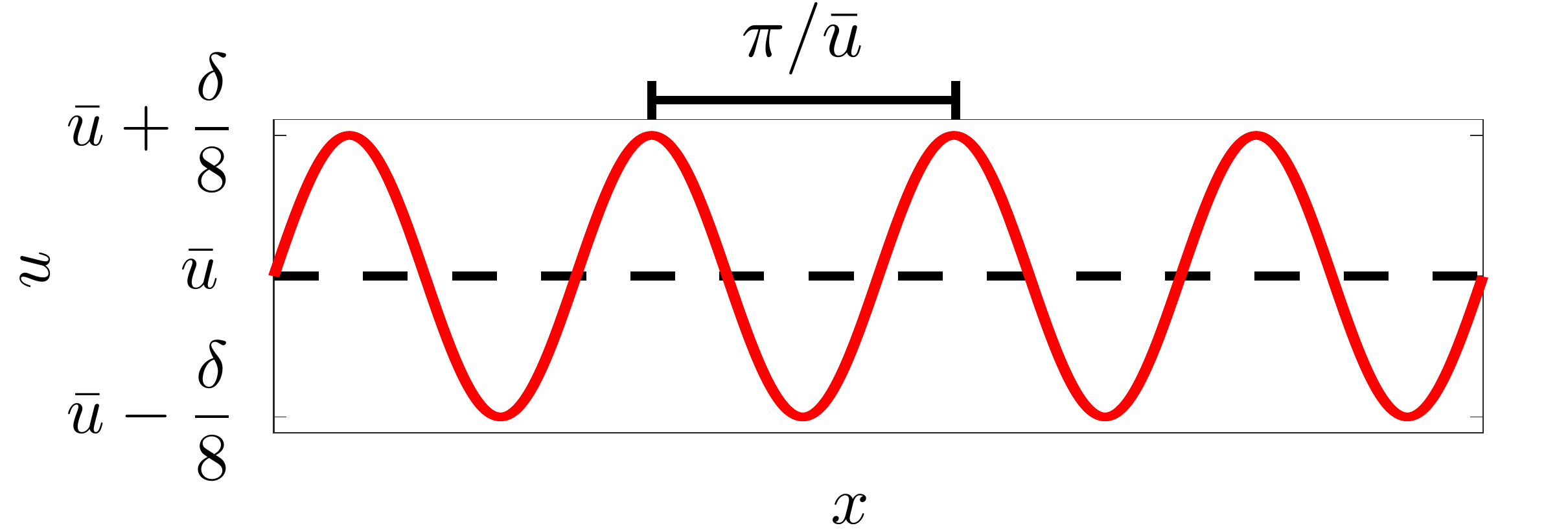}
\end{center}
\caption{\label{fig:SDW} Fluid velocity for an exact (dashed black line) and modulated (solid red line) SDW. }
\end{figure}

Comparing with the zero-thickness SDW, Eq.~\eqref{eq:SDW}, the only contribution of the nonlocal field in this approximation is a periodic perturbation in the fluid velocity with periodicity $2\ubar$. From a hydrodynamic perspective, this is a significant change since the constant fluid velocity becomes modulated at finite thicknesses, as shown in Fig.~\ref{fig:SDW}. In fact, we show below that the nonlocal dipole field correction leads to the appearance of a magnonic band structure.

\section{Magnon dispersion}

\subsection{Analytical calculation}

Here, we calculate the magnon dispersion for an approximate SDW, Eqs.~\eqref{eq:sdw_pert}. Because the fluid velocity is modulated by a weak perturbation proportional to $\delta$, Bloch's theorem can be invoked to calculate the dispersion relation to first order in $\delta$. For this, we assume a Bloch wave of the form $f(x)e^{i\mathbf{q}\mathbf{r}-i\omega t}$, where $f(x+N\pi/\bar{u})=f(x)$, $\omega$ and $\q = [q_x,q_y]$ are, respectively, the plane wave angular frequency and wavevector, $N$ is an integer, and $\mathbf{r}=\hat{\mathbf{x}} + \hat{\mathbf{y}}$. Because of the periodicity of Bloch waves, the resulting dispersion relation is that obtained for a zero-thickness SDW centered at every periodic point in reciprocal space, or Brillouin zones~\cite{Balkanski2000}. Following the procedure outlined in Refs.~\cite{Iacocca2017,Iacocca2017b}, we obtain
\begin{equation}\label{eq:disp}
\omega_\pm = \pm |\q+2N\ubar\hat{\mathbf{q}}_x|\sqrt{(1-\ubar^2) + |\q+2N\ubar\hat{\mathbf{q}}_x|^2},
\end{equation}
where $\hat{\mathbf{q}}_x=\left(\mathbf{q}\cdot\hat{\mathbf{x}}\right)\hat{\mathbf{x}}$. The $\pm$ sign in the dispersion denotes that waves can propagate parallel and anti-parallel to the wavevector. The spin waves associated with the dispersion \eqref{eq:disp} are fundamentally different from the traditional spin wave excitations observed in a uniformly magnetized film~\cite{Stancil2009}. We stress that we have assumed a homogeneous profile of the magnetization and plane waves in the film thickness, which is justified for thin films with $\delta<1$.

\subsection{Micromagnetic simulations}

The two-dimensional analytical magnon dispersion~\eqref{eq:disp} is validated numerically by micromagnetic simulations performed with the open-source code MuMax3~\cite{Vansteenkiste2014}. We utilize material parameters for Permalloy: saturation magnetization $M_s=790$~kA/m and exchange constant $A=10$~pJ/m, resulting in the exchange length $\lambda_\mathrm{ex}\approx5$~nm. We neglect the in-plane anisotropy so that the only source of symmetry-breaking is the nonlocal dipole field.

\begin{figure}
\begin{center}
\includegraphics[scale=0.44]{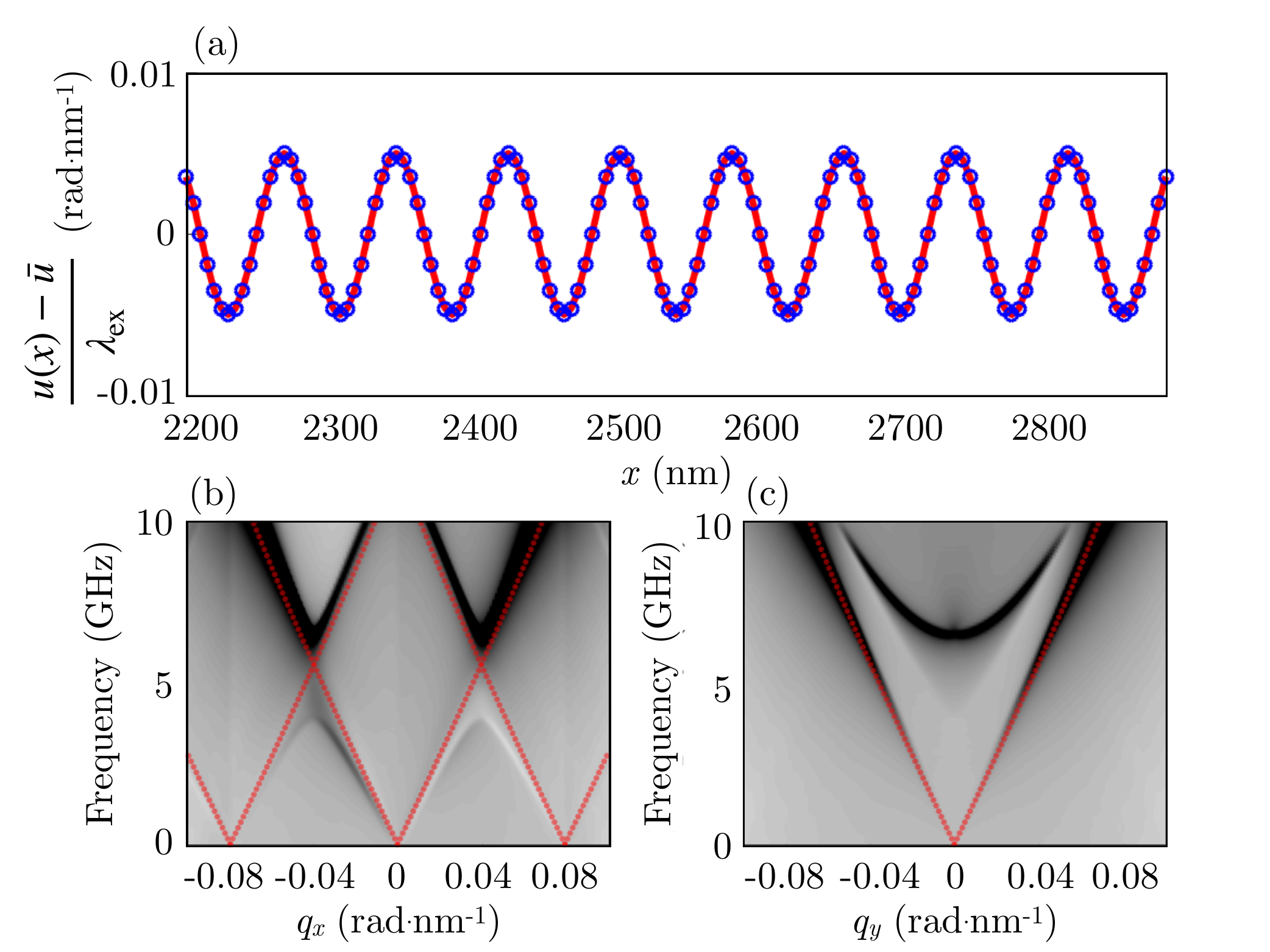}
\end{center}
\caption{\label{fig:groundstate} (a) Modulated fluid velocity for a Py thin film with thickness $d=1$~nm hosting a SDW with $\bar{u}/\lambda_\mathrm{ex}=0.4$~rad$\cdot$nm$^{-1}$. The micromagnetic results (blue circles) are in excellent agreement with the analytical approximation (solid red line). Cuts of the micromagnetically computed dispersion relation (contour plot) are shown for (b) $q_x$ with $q_y=0$ and (c) $q_y$ with $q_x=0$. Good agreement is obtained with the analytical dispersion relation (red dashed lines) considering Bloch waves along the SDW fluid velocity direction. }
\end{figure}

First, we numerically stabilize a SDW subject to nonlocal dipole field. For this computation, we define a thin film with a lateral side equal to an integer number of SDW periods and set periodic boundary conditions. The simulation is initialized with a zero-thickness SDW, Eq.~\eqref{eq:SDW} and is allowed to relax with high damping until equilibrium is obtained. To obtain enough spectral resolution for long waves, we discretize the simulation domain in $1024\times1024\times1$ cells. As a representative example, we use a SDW with $\bar{u}/\lambda_\mathrm{ex}=0.04$~rad$\cdot$nm$^{-1}$ and $d=1$~nm and obtain a thin film of size $5$~$\mu$m $\times~5$~$\mu$m $\times~1$~nm with cell-size $4.9$~nm $\times~4.9$~nm $\times~1$~nm. The fluid velocity of the numerically relaxed SDW can be calculated using Eq.~\eqref{eq:transf_u}. The numerical results are shown by blue circles in Fig.~\ref{fig:groundstate}(a). Excellent agreement with the approximate solution Eq.~\eqref{eq:uasymp} is found, as evidenced by the solid red curves. 

The magnon dispersion is numerically obtained by applying a space and time dependent field given by~\cite{Verkat2013}
\begin{eqnarray}\label{eq:field}
  \mu_0\mathbf{H}_\mathrm{ext} &=& B\mathrm{sinc}\left[q_{c,x}(x-x_0)\right]\mathrm{sinc}\left[q_{c,y}(y-y_0)\right]\nonumber\\ &&\times\mathrm{sinc}\left[2\pi f_c(t-t_0)\right]\hat{\mathbf{z}},
\end{eqnarray}
where we set $B=0.5$~T, $q_{c,x}=q_{c,y}=0.16$~rad$\cdot$nm$^{-1}$, $f_c=200$~GHz, $x_0=y_0=1.27$~$\mu$m, and $t_0=5$~ns. These parameters were set to optimize the spectral resolution within the first Brillouin zone (FBZ) $-\bar{u}/\lambda_\mathrm{ex}\leq q_x \leq\bar{u}/\lambda_\mathrm{ex}$. The simulation runs for $10$~ns. The magnon dispersion can be directly obtained by Fourier transformation of the resulting space and time dependent magnetization and no window functions are necessary due to the periodic boundary conditions. Cuts to the resulting magnon dispersion are shown in Fig.~\ref{fig:groundstate}(b) for the direction along the SDW fluid velocity, $q_x$ and $q_y=0$ and (c) for the direction perpendicular to the SDW fluid velocity, $q_y$ with $q_x=0$. Again, good quantitative agreement is observed with the analytical dispersion relation, Eq.~\ref{eq:disp}, upom time and space rescaling by Eq.~\eqref{eq:nondim}. Notably, Bloch waves are only established along the $\hat{\mathbf{q}}_x$ direction. Additionally, we observe that degeneracies in the dispersion are resolved by band gaps, as is typically the case in super-lattices~\cite{Balkanski2000}. The parabolic band above $\approx5$~GHz in Fig.~\ref{fig:groundstate}(c) is the result of aliased high-frequency projections along $\hat{\mathbf{q}}_y$ of the two-dimensional dispersion.

\begin{figure}
\begin{center}
\includegraphics[scale=0.4]{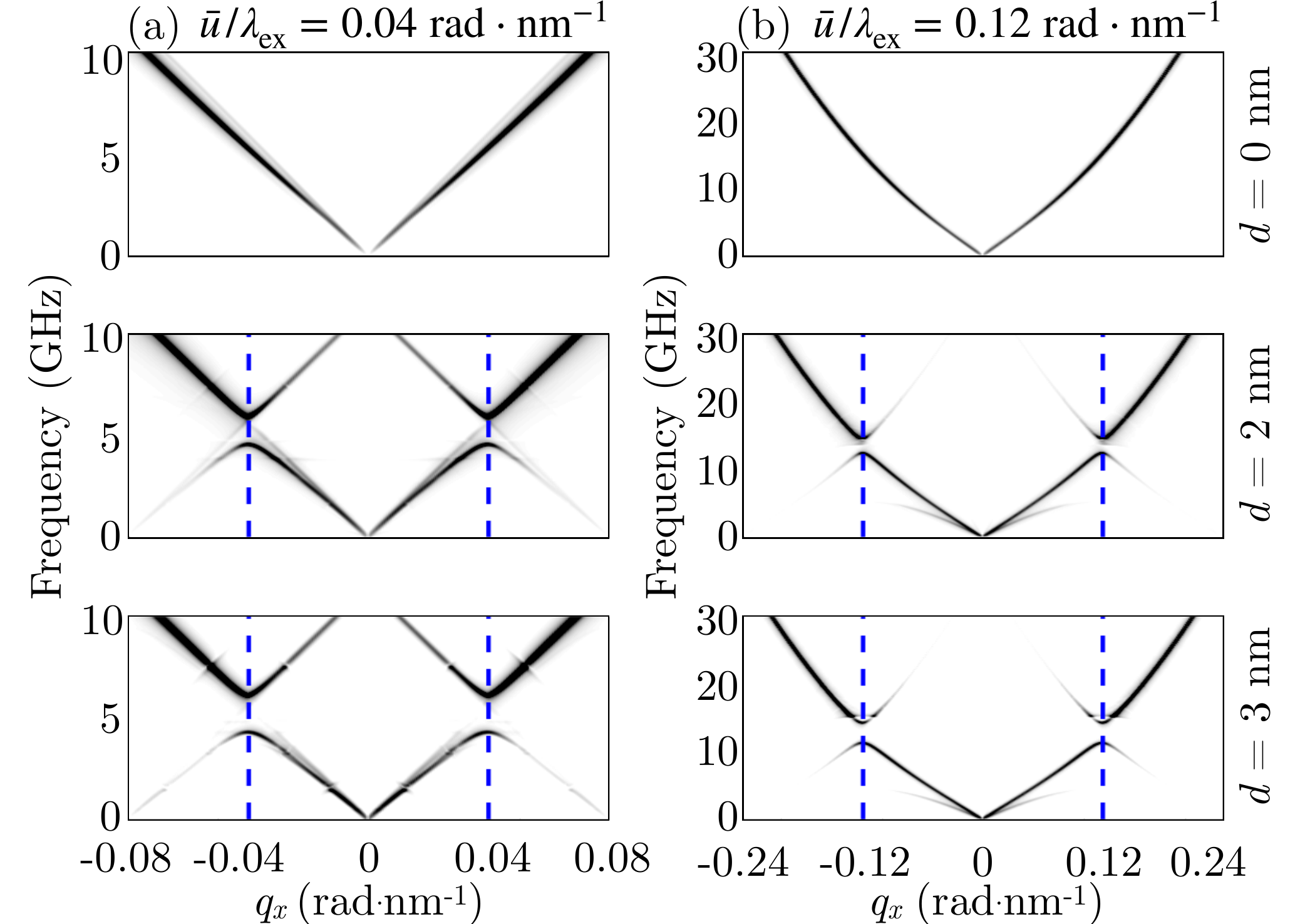}
\end{center}
\caption{\label{fig:results} Magnonic band structure on SDWs with (a) $\bar{u}/\lambda_\mathrm{ex}=0.04$~rad$\cdot$nm$^{-1}$ and $\bar{u}/\lambda_\mathrm{ex}=0.12$~rad$\cdot$nm$^{-1}$ obtained from micromagnetic simulations. The top row shows computations for an effectively zero thickness film. The middle and bottom row are obtained for $d=2$~nm and $d=3$~nm. For finite thicknesses, a band structure is observed accompanied by the emergence of band gaps at the FBZ, indicated by vertical blue dashed lines. }
\end{figure}

To investigate the band gaps in more detail, we restrict our dispersion calculation to the $\hat{\mathbf{q}}_x$ direction. For this, we define nanowires so that the simulation domain is discretized with $1024\times8\times1$ cells. As previously mentioned, the exact dimension of the nanowire coincides with an integer number of SDW periods. Simulations are performed by initializing zero-thickness SDWs with variable $\bar{u}/\lambda_\mathrm{ex}$, ranging from $0.04$~rad$\cdot$nm$^{-1}$ to $0.12$~rad$\cdot$nm$^{-1}$ in steps of $0.02$~rad$\cdot$nm$^{-1}$. For each initialized SDW, we first allow to relax the simulation for variable thicknesses ranging from 1~nm to 5~nm in steps of 1~nm. The resulting magnon dispersion is obtained by the method described earlier and setting $q_{c,y}=0$ in Eq.~\eqref{eq:field} to reduce wave reflections. The appearance of band structure as a function of thickness is shown in Fig.~\ref{fig:results}(a) and (b) for representative examples of SDWs with $\bar{u}/\lambda_\mathrm{ex}=0.04$~rad$\cdot$nm$^{-1}$ and $\bar{u}/\lambda_\mathrm{ex}=0.12$~rad$\cdot$nm$^{-1}$, respectively. The top row corresponding to $d=0$ was obtained by disabling the nonlocal dipole field calculation in MuMax3 and setting an effective perpendicular magnetic anisotropy constant of $K_u=-\mu_0M_s^2/2$~J/m$^3$ corresponding to the local dipole field. When a finite thin film thickness is considered, shown in the subsequent rows, we observe the emerging band structure and band gaps at the FBZ, $q = \pm \bar{u}/\lambda_\mathrm{ex}$, indicated by vertical dashed blue lines.

\begin{figure}
\begin{center}
\includegraphics[scale=0.48]{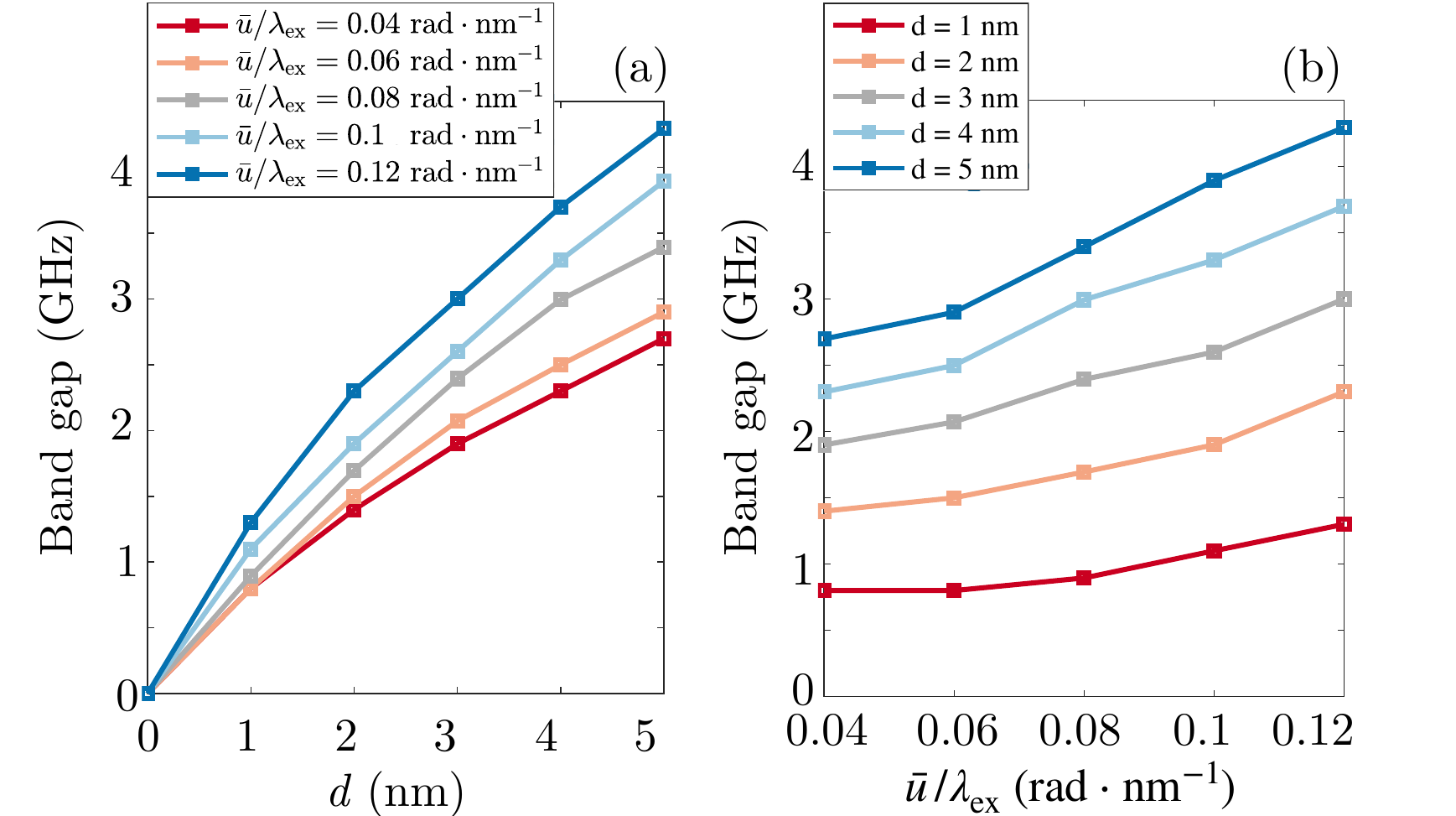}
\end{center}
\caption{ \label{fig:BG_params} Measured band gap at the FBZ as a function of (a) $d$ and (b) $\ubar/\lambda_\mathrm{ex}$ from micromagnetic simulations.}
\end{figure}

From the numerically computed band structures, we measure the band gap at the FBZ as a function of both $d$ and $\ubar/\lambda_\mathrm{ex}$, shown in Fig.~\ref{fig:BG_params}. It is observed that the band gap increases with both parameters and is on the order of GHz. A sub-linear power law dependence is observed as a function of $d$, Fig.~\ref{fig:BG_params}(a). As a function of $\bar{u}/\lambda_\mathrm{ex}$, the band gap increases in a manner without a clear functional form.

\section{Conclusion}

In this letter, we have analytically and numerically studied the effect of the nonlocal dipole field on SDWs and the associated magnon dispersion in ferromagnetic thin films. Utilizing a hydrodynamic interpretation, it was found that the nonlocal dipole field modulates the fluid velocity at twice the SDW wavenumber. This periodicity induces a band structure in the magnon dispersion. Band gaps are numerically found for Py parameters on the order of GHz with a magnitude that depends on both the film thickness and the SDW wavenumber. Consequently, the magnonic band structure can be reconfigured by changing the SDW wavenumber, e.g., by spin injection at the extrema of ferromagnetic nanowires.

\section*{Acknowledgment}
This material is based on work supported by the U.S. Department of Energy, Office of Science, Office of Basic Energy Sciences under Award Number 0000231415. 




\bibliographystyle{IEEEtran}
%


\end{document}